\documentclass[preprint]{iucr}

\usepackage[latin1]{inputenc}
\usepackage{amssymb}
\usepackage{amsmath}
\usepackage{amsthm}
\usepackage{bmpsize}
\usepackage{color}
\usepackage[english]{babel}
\usepackage{exscale}
\usepackage{graphics}
\usepackage{graphicx}
\usepackage{harvard}
\usepackage{longtable}
\usepackage{mathtools}
\usepackage{pdflscape}
\usepackage{relsize}
\usepackage{soul}
\usepackage{tabularx}

\begin{document}

\title{Superstructure reflexions in tilted perovskites}
\shorttitle{Octahedral tilting rules}

\author[a]{Richard}{Beanland}
\cauthor[a]{Robin}{Sj\"okvist}{Robin.Sjokvist@warwick.ac.uk}

\aff[a]{Department of Physics, University of Warwick, \city{Coventry} CV4 7AL, \country{UK}}

\maketitle

\begin{abstract}
The superstructure spots that appear in diffraction patterns of tilted perovskites are well documented and easily calculated using crystallographic software.  Here, by considering a distortion mode as a perturbation of the prototype perovskite structure, we show how the structure factor equation yields Boolean conditions for the presence of first order superstructure reflexions.  A subsequent article describes conditions for second order reflexions, which appear only in structures with mixed in-phase and anti-phase oxygen octahedral tilting.  This approach may have some advantages for the analysis of electron diffraction patterns of perovskites.
\end{abstract}

\section{Introduction}

~

Conditions for the appearance of superstructure reflexions in diffraction from $ABO_3$ perovskites with $BO_6$ octahedral tilting were outlined almost exactly fifty years ago in the seminal work of Glazer \cite{Glazer72, Glazer75}.  In X-ray and neutron diffraction, where many diffracted beams are routinely collected over a wide range of crystal orientations, these reflexions, and changes in unit cell dimensions, may be used to determine the space group, extinction conditions and thus the crystal structure.  However in transmission electron microscopy (TEM) and electron diffraction (ED), excluding 3D-ED methods \cite{Gemmi} it is usual to examine a few low index zone axes in individual crystals or domains, which provides only partial data.  Nevertheless, this can provide crucial information that is sufficient to distinguish between alternative structures \cite{Woodward}. In addition, the sensitivity of electron scattering to low atomic number (i.e. oxygen) atoms and the ability to probe nanoscale regions gives ED an important role in the characterisation of perovskite oxides.

With incomplete knowledge of the three-dimensional reciprocal lattice and space group, in ED it is common to work in the reference frame of the prototype perovskite structure while using the term `pseudo-cubic' to acknowledge that this is not actually a correct description of the structure.  In the pseudo-cubic reference frame, superstructure reflexions that result from larger periods in direct space appear at fractional coordinates in reciprocal space, i.e. doubled lattice translations produced by octahedral tilting give reflexions at half-order positions.  The different patterns of superstructure reflexions produced by different Glazer tilt systems has been determined by inspecting simulations \cite{Woodward} for some low index ED patterns.  Here, we revisit this question and derive general equations for superstructure reflexions.  The emphasis on tilt system (or other distortion mode) rather than space group, avoids the need to change reference frame according to different choices of unit cell and the conversion of Miller indices describing the zone axis, reciprocal lattice vectors, and systematic absences this entails. The approach is therefore convenient when analysing diffraction patterns of perovskites exhibiting different distortion modes, as well as providing a result for any zone axis.

\section{Calculation}

~

Ignoring thermal factors, the structure factor equation that gives the complex amplitude of a diffracted beam $\textbf{g}$ in a crystal with a static distortion mode can be written

\begin{equation}
    \label{Struc}
    F_g = \sum_{j=1}^{n} f_g^{(j)} \exp{\left(2 \pi i \textbf{g} \cdot (\textbf{r}^{(j)} + \boldsymbol{\delta}^{(j)})\right)},
\end{equation}

where the sum is taken over all $j$ atoms in the unit cell, each having atomic scattering factor $f_g^{(j)}$, fractional coordinates $\textbf{r}^{(j)}$ in the prototype structure, and static displacement from these prototype coordinates (due to a distortion mode, such as an oxygen octahedral tilt system) $\boldsymbol{\delta}^{(j)}$.

Here, we are not interested in the precise value of $F_g$ for a superstructure reflection. Rather, our main concern is whether a distortion mode produces a superstructure reflexion, or not.  The answer to this question is simply that a reflexion will be present when the result of Eq.~\ref{Struc} is not exactly zero and, as is shown below, this can be determined most easily by allowing $\boldsymbol{\delta}^{(j)}$ to be arbitrarily small. This approach also means that any second order effects (e.g. octahedral distortions) can be neglected. Using the approximation $e^{(a+b)} = e^a e^b = e^a (1+b)$ for small $b$, and noting that the structure factor for superstructure reflexions in the prototype structure is precisely zero, the structure factor of a superstructure reflexion with infinitesimal $\boldsymbol{\delta}^{(j)}$ is

\begin{equation}
    \label{SuperStruc}
    F_g = \sum_{j=1}^{n} f_g^{(j)} 2 \pi i \textbf{g} \cdot \boldsymbol{\delta}^{(j)} \exp{\left(2 \pi i \textbf{g} \cdot \textbf{r}^{(j)}\right)},
\end{equation}

On first sight Eq.~\ref{SuperStruc} does not appear to be much more informative than Eq.~\ref{Struc}, but further simplification can be obtained, as follows.

We choose a unit cell that is twice the size of the prototype in all three dimensions, which is large enough to be a unit cell for any Glazer octahedral tilting pattern (although it will not generally correspond to the fundamental unit cell of the distorted structure).  In this reference frame, superstructure reflexions have integer Miller indices with at least one odd index, while reflexions of the prototype structure have all-even indices (i.e. integer indices in the pseudo-cubic frame).  This expanded cell is eight times the size of the prototype cell and has 24 oxygen atoms with coordinates given in Table~\ref{oxy}.  Here, for reasons that will shortly become apparent we write the oxygen coordinates $r_i$, which all have positions that are multiples of a quarter of the lattice parameter of the unit cell, with an integer form $s_i = 4r_i$.  The exponential term can then be written

\begin{equation}
    \label{gr_nice}
    \begin{split}
    &\exp{(2 \pi i \textbf{g} \cdot \textbf{r}^{(j)})} \\
    &= \exp{(s_1 g_1 \pi i/2)} \exp{(s_2 g_2 \pi i/2)} \exp{(s_3 g_3 \pi i/2)} \\
    &= A^{s_1}B^{s_2}C^{s_3}
    \end{split}
\end{equation}

using the substitution $A=\exp{(g_1 \pi i/2)}$, $B=\exp{(g_2 \pi i/2)}$ and $C=\exp{(g_3 \pi i/2)}$.  The Miller indices $g_i$ are integers and thus the terms $A$, $B$ and $C$ take values of $\pm 1$ for even $g_i$ and $\pm i$ for odd $g_i$.

\begin{table}
    \caption{Coordinates of the 24 oxygen atoms in the expanded unit cell used to describe tilted perovskites (Fig.~\ref{deltapic}, written as integer multiples of 1/4, e.g. $\textbf{r}_{O1}~=~[0,1/4,1/4]$.}
    \label{oxy}
    \begin{center}
        \begin{tabular}{c|c c c c c|c c c}
            Atom & $s_1$ & $s_2$ & $s_3$ & & Atom & $s_1$ & $s_2$ & $s_3$ \\
            \hline
            O1 & 1 & 0 & 1 & & O13 & 1 & 2 & 1 \\
            O2 & 1 & 0 & 3 & & O14 & 1 & 2 & 3 \\
            O3 & 3 & 0 & 1 & & O15 & 3 & 2 & 1 \\
            O4 & 3 & 0 & 3 & & O16 & 3 & 2 & 3 \\
            O5 & 0 & 1 & 1 & & O17 & 0 & 3 & 1 \\
            O6 & 0 & 1 & 3 & & O18 & 0 & 3 & 3 \\
            O7 & 1 & 1 & 0 & & O19 & 1 & 3 & 0 \\
            O8 & 1 & 1 & 2 & & O20 & 3 & 3 & 2 \\
            O9 & 2 & 1 & 1 & & O21 & 2 & 3 & 1 \\
            O10 & 2 & 1 & 3 & & O22 & 2 & 3 & 3 \\
            O11 & 3 & 1 & 0 & & O23 & 3 & 3 & 0 \\
            O12 & 3 & 1 & 2 & & O24 & 3 & 3 & 2
        \end{tabular}
    \end{center}
\end{table}

\begin{figure}
    \caption{Displacements of oxygen atoms due to a $BO_6$ octahedral rotation about the $c$-axis.  Labels correspond to Table~\ref{oxy}; red arrows indicate displacements $\boldsymbol{\delta}^{(j)}$ listed in Table~\ref{c+}.}
    \label{deltapic}
    \centering \includegraphics[width=0.6\columnwidth]{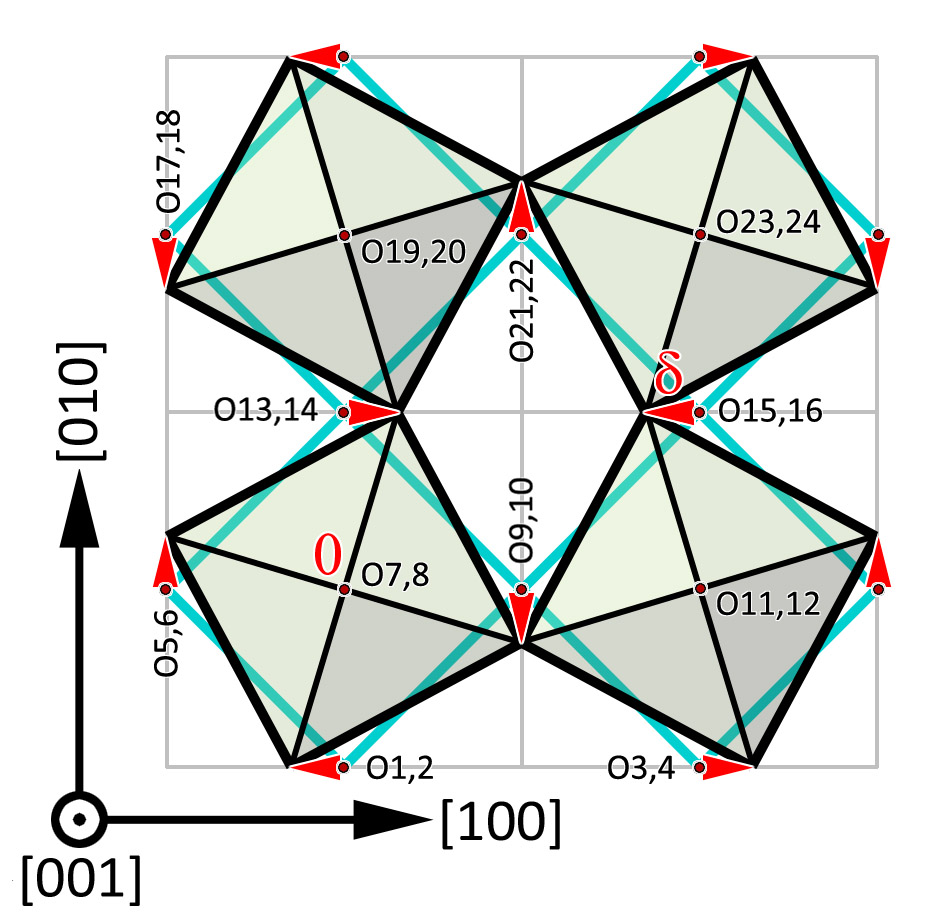}
\end{figure}

We are now ready to consider specific distortion modes.  Fig.~\ref{deltapic} shows the direction of (infinitesimal) oxygen atom displacements for in-phase octahedral rotations about $c$, $a^0 a^0 c^+$ in Glazer notation, which are listed in Table~\ref{c+}.  Using Eq.~\ref{gr_nice} and substituting into  Eq.~\ref{SuperStruc} we obtain

\begin{equation}
    \label{polysum}
    \begin{split}
    \frac{F_g}{2 \pi i f_g \delta} = & -g_1 AC -g_1 AC^3 +g_1 A^3C +g_1 A^3C^3 \\
    & +g_2 BC +g_2 BC^3 -g_2 A^2BC -g_2 A^2BC^3 \\
    & +g_1 AB^2C +g_1 AB^2C^3 -g_1 A^3B^2C -g_1 A^3B^2C^3 \\
    & -g_2 B^3C -g_2 B^3C^3 +g_2 A^2B^3C +g_2 A^2B^3C^3,
    \end{split}
\end{equation}

which nicely reduces to

\begin{equation}
    \label{SSc+}
    \frac{F_g}{2 \pi i f_g \delta} = C \left(g_2 B - g_1 A \right)
    \left(1 - A^2\right)\left(1 - B^2\right)\left(1 + C^2 \right).
\end{equation}

This equation can be interpreted as a set of Boolean conditions, all of which must be satisfied for a superstructure reflection to exist.  Thus, since $A=\pm 1$ for even $g_1$ and $A=\pm i$ for odd $g_1$, $(1 - A^2)$ is only non-zero, and a superstructure reflexion will only be present, when the first index of the reflexion, $g_1$, is odd.  Similarly, $(1 + C^2)$ is only non-zero for even $g_3$ and therefore Eq.~\ref{SSc+} indicates that superstructure reflexions of the $a^0 a^0 c^+$ tilt system must have the form odd-odd-even in the frame of the doubled cell, which can be written in shorthand using the prototype cell as $\frac{1}{2}ooe$.  As for the other two terms in Eq.~\ref{SSc+},  $C=\exp{(g_3 \pi i/2)}$ is never zero, while $(g_2 B - g_1 A) = 0$ when $|g_1|=|g_2|$.  We thus obtain the result that superstructure reflexions occur with pseudo-cubic indices $\frac{1}{2}ooe$, $|g_1| \ne |g_2|$. The latter condition describes the systematic absences that result from the $b$-glide plane in the space group of the $P4/mbm$ $a^0 a^0 c^+$ structure.

\begin{table}
    \caption{Oxygen atom displacements produced by the $a^0a^0c^+$ tilt system shown in Fig.~\ref{deltapic}.  Displacements in the $a^0a^0c^-$ tilt system are similar, except O2, O4, O6, O10, O14, O16, O18 and O22 whose displacements are reversed. }
    \label{c+}
    \begin{center}
        \begin{tabular}{c|c c c c c|c c c}
            Atom & $\delta_1$ & $\delta_2$ & $\delta_3$ & & 
            Atom & $\delta_1$ & $\delta_2$ & $\delta_3$ \\
            \hline
            O1 & $-\delta$ & 0 & 0 & & O13 & $\delta$ & 0 & 0 \\
            O2 & $-\delta$ & 0 & 0 & & O14 & $\delta$ & 0 & 0 \\
            O3 & $\delta$ & 0 & 0 & & O15 & $-\delta$ & 0 & 0 \\
            O4 & $\delta$ & 0 & 0 & & O16 & $-\delta$ & 0 & 0 \\
            O5 & 0 & $\delta$ & 0 & & O17 & 0 & $-\delta$ & 0 \\
            O6 & 0 & $\delta$ & 0 & & O18 & 0 & $-\delta$ & 0 \\
            O7 & 0 & 0 & 0 & & O19 & 0 & 0 & 0 \\
            O8 & 0 & 0 & 0 & & O20 & 0 & 0 & 0 \\
            O9 & 0 & $-\delta$ & 0 & & O21 & 0 & $\delta$ & 0 \\
            O10 & 0 & $-\delta$ & 0 & & O22 & 0 & $\delta$ & 0 \\
            O11 & 0 & 0 & 0 & & O23 & 0 & 0 & 0 \\
            O12 & 0 & 0 & 0 & & O24 & 0 & 0 & 0 \\
        \end{tabular}
    \end{center}
\end{table}





A similar procedure can be performed for the $a^0a^0c^-$ tilt system, in which the displacements of oxygen atoms with even-numbered labels in Table~\ref{c+} are reversed, with the result

\begin{equation}
    \label{SSc-}
    \frac{F_g}{2 \pi i f_g \delta} = C \left(g_2 B - g_1 A \right)
    \left(1 - A^2\right)\left(1 - B^2\right)\left(1 - C^2 \right),
\end{equation}

indicating that superstructure reflections must have pseudo-cubic form $\frac{1}{2}ooo$ with the same systematic absences, this time from the $c$-glide in the space group $I4/mcm$ of the $a^0a^0c^-$ structure.

Other distortion modes, such as antiferrodistortive cation displacements, can also be considered in a similar manner using their coordinates and displacements.

A particularly elegant aspect of this approach is that rules for superstructure reflections in structures with oxygen octahedral tilts about multiple axes -- or, indeed, multiple distortion modes (e.g. antiferrodistortive displacements, distorted oxygen octahedra etc.) --can be constructed simply by adding equations, giving a straightforward and quick method of calculation.  The results are summarised for the 14 crystallographically distinct oxygen octahedral tilt systems \cite{H+S} in the appendix.

\section{Conclusions}

~

Equations governing the appearance of superstructure spots resulting from distortion modes in perovskites have been derived.  This may aid the interpretation of electron diffraction patterns and replicates the work of \cite{Glazer75} and \cite{Woodward}.  The emphasis on distortion mode, rather than space group, allows the interpretation of ED patterns without the need to rewrite vectors in real and reciprocal space for different unit cells.

\section{Acknowledgements}

~

We thank Prof. A. M. Glazer for constructive comments.  This work was funded by EPSRC grant EP/V053701/1.

\newpage

\section{\textbf{Appendix}}

~

For completeness we compile in Table~\ref{summary} the conditions governing the existence of superstructure spots in the pseudo-cubic reference frame for the 14 crystallographically distinct Glazer tilt systems as listed by Howard and Stokes \cite{H+S}.  These may be obtained by adding equations similar to \ref{SSc+} and/or \ref{SSc-} for the appropriate tilt system.

~

\begin{table}
    \caption{Pseudo-cubic superstructure reflections in perovskites with octahedral tilting. Miller indices are given in the form $\textbf{g}=hkl$ to allow for concise descriptions of extinctions}
    \label{summary}
        \begin{tabular}{m{0.05\linewidth} | m{0.1\linewidth} | m{0.1\linewidth} | m{0.75\linewidth}}
            No. & Tilt \mbox{system} & Space group & Conditions for superstructure spots to exist\\
            \hline
            1 &\(a^+a^+a^+\) & \(Im\bar{3}\) &
            \begin{tabular}[c]{@{}l@{}}\(\frac{1}{2}ooe,~|h| \neq |k|;~\frac{1}{2}oeo,~|h| \neq |l|;\)\\\(\frac{1}{2}eoo,~|k| \neq |l| \)\end{tabular} \\
            ~ & ~ & ~\\
            2 & \(a^0b^+b^+\) & \(I\frac{4}{m}mm\) & \(\frac{1}{2}ooe,~|h| \neq |k|;~\frac{1}{2}oeo,~|h| \neq |l| \) \\
            ~ & ~ & ~\\
            3 &\(a^0a^0c^+\) & \(P\frac{4}{m}bm\) & \(\frac{1}{2}ooe,~|h| \neq |k| \) \\
            ~ & ~ & ~\\
            4 & \(a^0a^0c^-\) & \(I\frac{4}{m}cm\) & \(\frac{1}{2}ooo,~|h| \neq |k| \) \\
            ~ & ~ & ~\\
            5 & \(a^0b^-b^-\) & \(Imma\) & \begin{tabular}[c]{@{}l@{}}\(\frac{1}{2}ooo,~!(|h|=|k|=|l|),~k \ne l, \)\\!\((h=n, k=\pm n+4m, l=\mp n+4m),\)\\ \(n, m\) integers\end{tabular} \\
            ~ & ~ & ~\\
            6 & \(a^-a^-a^-\) & \(R\bar{3}c\) & \begin{tabular}[c]{@{}l@{}}\(\frac{1}{2}ooo,~h \ne k~\&~k \ne l~\&~l \ne h, \)\\!\((h=n, k=n+4m, l=n+4p),\)\\ \(n, m, p\) integers\end{tabular} \\
            ~ & ~ & ~\\
            7 & \(a^+b^+c^+\) & \(Immm\) & \begin{tabular}[c]{@{}l@{}}\(\frac{1}{2}ooe,~|h| \neq |k|;~\frac{1}{2}oeo,~|h| \neq |l|;\)\\\(\frac{1}{2}eoo,~|k| \neq |l| \)\end{tabular} \\
            ~ & ~ & ~\\
            8 & \(a^+a^+c^-\) & \(P\frac{4_2}{n}mc\) & \begin{tabular}[c]{@{}l@{}}\(\frac{1}{2}ooo,~|h| \neq ~|k|;~\frac{1}{2}oeo,~|h| \neq ~|l|;\)\\\(\frac{1}{2}eoo,~|k| \neq ~|l| \)\end{tabular} \\
            ~ & ~ & ~\\
            9 & \(a^0b^+c^-\) & \(Cmcm\) & \(\frac{1}{2}ooo,~|h| \neq |k|;~\frac{1}{2}oeo,~|h| \neq |l| \) \\
            ~ & ~ & ~\\
            10 & \(a^+b^-b^-\) & \(Pnma\) & \begin{tabular}[c]{@{}l@{}}\(\frac{1}{2}eoo,~|k| \neq |l|;\)\\\(\frac{1}{2}ooo,~!(|h|=|k|=|l|),~k \ne l, \)\\!\((h=n, k=\pm n+4m, l=\mp n+4m),\)\\ \(n, m\) integers\end{tabular} \\
            ~ & ~ & ~\\
            11 & \(a^0b^-c^-\) & \(C\frac{2}{m}\) & \(\frac{1}{2}ooo, ~!(|h|=|k|=|l|)\) \\
            ~ & ~ & ~\\
            12 & \(a^-b^-b^-\) & \(C\frac{2}{c}\) & \(\frac{1}{2}ooo, ~!(|h|=|k|=|l|), ~k \neq l\) \\
            ~ & ~ & ~\\
            13 & \(a^+b^-c^-\) & \(P\frac{2_1}{m}\) & \(\frac{1}{2}ooo, ~!(|h|=|k|=|l|);~\frac{1}{2}eoo,~|k| \neq |l| \) \\
            ~ & ~ & ~\\
            14 & \(a^-b^-c^-\) & \(P\bar{1}\) & \(\frac{1}{2}ooo, ~!(|h|=|k|=|l|)\) \\
        \end{tabular}
\end{table}

Some general rules become apparent from Table~\ref{summary}.  The rules for in-phase tilting are quite straightforward, with each tilt system $a^+$, $b^+$, $c^+$ producing its own set of superstructure spots with pseudo-cubic indices $\frac{1}{2}eoo$, $\frac{1}{2}oeo$, $\frac{1}{2}ooe$ with no dependence on the presence of any other distortions.  Conversely, all antiphase tilt systems produce pseudo-cubic $\frac{1}{2}ooo$ superstructure spots, and are only distinguished by their systematic absences. Furthermore, because in these cases the type of superstructure reflexions is always the same, tilts of equal magnitude operating about different axes can result in changes to the set of systematic absences.  Accordingly, systematic absences are most apparent for the $a^-a^-a^-$ system.  This means that determining antiphase tilting systems is less straightforward than in-phase tilt systems.  For investigations using electron diffraction, it may thus be important to explore reciprocal space in three dimensions since systematic absences can readily be `filled in' by double diffraction where the possibility exists, particularly in the zero-order Laue zone.  Access to higher order Laue zones, or zone axes where no double diffraction pathways are present, is generally necessary.

Table~\ref{summary} shows that reflexions of the form $\frac{1}{2}eeo$, $\frac{1}{2}eoe$ and $\frac{1}{2}oee$ do not result from oxygen octahedral tilting.  They may, however, be produced by antiferrodistortive displacements of cations.  Calculation of extinction rules for these distortion modes is left as an exercise for the reader.

\referencelist[references]


@article{H+S,
author = {Howard, C. J. and Stokes, H. T.},
title = {Group-Theoretical Analysis of Octahedral Tilting in Perovskites},
journal = {Acta Crystallographica Section B},
year = {1998},
volume = {54},
number = {6},
pages = {782--789},
month = {Dec}
}

@article{Gemmi,
    author = {Gemmi, Mauro and Mugnaioli, Enrico and Gorelik, Tatiana and Kolb, U. and Palatinus, Luk{\'{a}}{\v{s}} and Boullay, Philippe and Abrahams, Jan},
    title = {3D Electron Diffraction: The Nanocrystallography Revolution},
    journal = {ACS Central Science},
    year = {2019},
    volume = {5},
    pages = {1315--1329},
    doi = {10.1021/acscentsci.9b00394}
}

@article{Glazer72,
   author = {Glazer, A. M.},
   title = {The classification of tilted octahedra in perovskites},
   journal = {Acta Crystallographica Section B},
   volume = {28},
   number = {11},
   pages = {3384-3392},
   ISSN = {0567-7408},
   DOI = {doi:10.1107/S0567740872007976},
   year = {1972},
   type = {Journal Article}
}

@article{Glazer75,
author = {Glazer, A. M.},
title = {Simple ways of determining perovskite structures},
journal = {Acta Crystallographica Section A},
volume = {31},
number = {6},
pages = {756-762},
doi = {https://doi.org/10.1107/S0567739475001635},
year = {1975}
}

@article{Woodward,
author = {Woodward, David I. and Reaney, Ian M.},
title = {Electron diffraction of tilted perovskites},
journal = {Acta Crystallographica Section B},
year = {2005},
volume = {61},
number = {4},
pages = {387--399},
month = {Aug},
doi = {10.1107/S0108768105015521}
}
\end{document}